\documentclass[aps]{revtex4}
\def\be {\begin{equation}}
\def\ee {\end{equation}}
\def\ba {\begin{eqnarray}}
\def\ea {\end{eqnarray}}
\def\nn {\nonumber}
\def\a  {\alpha}
\def\b  {\beta}
\def\c  {\gamma}
\def\C  {\Gamma}
\def\d  {\delta}

\def\e  {\epsilon}

\def\l  {\lambda}

\def\m  {\mu}
\def\n  {\nu}
\def\o  {\omega}
\def\O  {\Omega}
\def\p  {\pi}

\def\r  {\rho}

\def\la {\label}
\def\le {\left}
\def\ri {\right}
\def\pa {\partial}
\def\f {\frac}

\def\no {\noindent}
\def\bi {\begin{itemize}}
\def\ei {\end{itemize}}

\def\ra {\rangle}
\def\vs {\vspace}

\def\ul {\underline}

\def\laq{\hbox{~}\raise 0.4ex\hbox{$<$}\kern -0.8em\lower 0.62ex\hbox{$\sim$}\hbox{~}}
\def\gaq{\hbox{~}\raise 0.4ex\hbox{$>$}\kern -0.7em\lower 0.62ex\hbox{$\sim$}\hbox{~}}

\begin{document}
\medskip
\hfill  UNB Technical Report 04-02\\[20pt]

\title{Gravitational non-commutativity and G\"odel-like spacetimes}

\author{Saurya Das}

\affiliation{Dept. of Physics,
University of Lethbridge, \\
4401 University Drive,
Lethbridge, Alberta - T1K 3M4, CANADA}

\author{Jack Gegenberg}

\affiliation{Dept. of Mathematics
and Statistics, \\ University of New Brunswick,  \\
Fredericton, New Brunswick - E3B 5A3, CANADA}

\begin{abstract}

We derive general conditions under which geodesics of stationary spacetimes
resemble trajectories of charged particles in an electromagnetic field.
For large curvatures (analogous to strong magnetic fields),
the quantum mechanicical states of these particles
are confined to gravitational analogs of {\it lowest Landau levels}.
Furthermore, there is an effective non-commutativity between their
spatial coordinates.
We point out that the Som-Raychaudhuri and G\"odel spacetime
and its generalisations are precisely of the
above type and compute the effective non-commutativities that they induce.
We show that the non-commutativity for G\"odel spacetime is identical
to that on the fuzzy sphere. Finally, we show how the star
product naturally emerges in Som-Raychaudhuri spacetimes.

\end{abstract}

\maketitle

\section{Introduction}

It is conceivable that the as-yet-undiscovered fundamental theory that
underlies (quantum) gravity and the interactions of particles will be
characterized by a non-commutative geometry.  Indeed, there is some
evidence in string theory that this is the case \cite{sw}. The origins
of this non-commutativity seem to be associated with strong magnetic-like
fields that occur in various string and supergravity theories \cite{mr}.  At a
seemingly less fundamental level, the quantum theory of a charged particle
in an external magnetic field - the Landau atom - is associated with an
effective spatial non-commutativity.

But if spacetime is in some sense non-commutative, then it should show
up in the structures that are collectively know as geometry.
Similarities
between the geodesic equation for gravity and the Lorentz force equation for
electromagnetism are well known. The analogy is further strengthened
for stationary spacetimes,
where not only is there an exact correspondence between
different components of the metric and electric and magnetic fields (such
that the geodesic equation can be exactly cast in the form of Lorentz force
equation), but that given certain special stationary spacetimes,
particles in them exhibit interesting phenemonena such as Landau levels
and spatial non-commutativity. One such class of spacetimes was
discovered by
Som and Raychaudhuri and later re-derived using simple matter field
configurations of electromagnetic and scalar fields by
one of us (JG) and A. Das \cite{somray,gd}. We will refer to them
as SR spacetimes.
We examine some aspects of these spacetimes and compute the
Landau levels as well as spatial non-commutativities that they induce.
Similarities between the Landau atom and the above spacetimes was
discovered by Hikida et al \cite{hikida} and Drukker et al \cite{drukker}.
Here we present an underlying
reason for this, and as well we further extend their results
and show that magnetic fields responsible for ordinary Landau levels
are naturally
associated with these spacetimes as their sources, and that
their spatial coordinates exhibit an effective non-commutativity.

What we obtain is the backbone of a cosmology with effective spatial
non-commutativity occuring at a large scale.  The scale is given by
the rotation parameter $\Omega$, which also can be interpreted as the
scale of a magnetic-like, or twist, gravitational field.  It also describes
the boundary of the `causally safe region' of the spacetime. That is in
the region where the radial coordinate satisfies $0<r < 1/\Omega$, there
are no closed timelike or null curves.  In the neighbourhood of $r=0$, the
spacetime appears Minkowskian.

We want to begin to explore the idea here that effective spatial
non-commutativity emerges from the behaviour
of geodesics in a realm where quantum theory is relevant.  We
believe that there is an analogy here to black hole thermodynamics.
In the latter the existence of analogies between the properties of
geometrical quantities
associated with black hole geometry and classical thermodynamics is
often interpreted as revealing underlying quantum field theoretic issues
associated with gravity.  In our case, what is revealed is that in the
case of very strong gravito-magnetic fields, effective spatial
non-commutativity can be observed in a sector of the quantum mechanics of
particles in that gravitational background.
Furthermore, such non-commutativity is closely related to that on the
fuzzy sphere, and naturally gives rise to the star product in
these spacetimes.

This article is organised as follows:
in the next section, we recast the
geodesic equations of a charged particle moving
in electromagnetic and stationary
gravitational fields in the form of a combined
Lorentz force-like equation, in which the
particle is subjected to `effective' eletromagnetic fields which include
contributions from spacetime curvature. In section (\ref{sec3})
we review literature where it was shown
that charges in strong magnetic fields
experience an effective non-commutativity of spatial coordinates, and extend
the results to include gravitational fields. That is, just as
electromagnetism, gravity can induce non-commutativity.
In section (\ref{sec4}), we discuss SR spacetime and its generalization as
G\" odel/AdS metrics as solutions of the
Einstein equations with appropriate sources.
In section (\ref{sec5}) we further show that for
strong enough curvatures of these metrics,
energetic particles are confined to analogs of
lowest Landau levels and that their spatial coordinates become
non-commutative.
In section (VI), we show that non-commutativity on G\"odel
spacetime is identical to that on the fuzzy sphere, and in
section (VII), we show that the star-product emerges for SR spacetimes.
Finally we conclude in section (\ref{sec6})
with a summary and some open questions.

\section{Stationary Spacetimes and Electromagnetism}

We start with the general case of
a particle of mass $m$ and charge $e$ in a curved spacetime with metric
$g_{\mu\nu}$
and electromagnetic field $F_{\m\n}$.  For the most part, we are interested
in the uncharged case, but we consider the charged case in order to
see most clearly the gravitational analogue of charged
particle motion in an external
magnetic field.

The action functional is
\be
S[x^0,x^a]:=\int^{\tau_2}_{\tau_1} d\tau\left[{mc^2\over2}g_{\m\n}\dot x^\m
\dot x^\n+{e\over c}A_\m\dot x^\m\right].\label{laction}
\ee
We write $\dot x^\m:=dx^\m/cd\tau$.  In the above, in general, the metric
$g_{\m\n}$ and 4-potential $A_\m$ depend on all the coordinates.
We assume that the metric and the electromagnetic fields are stationary.
That
is, there exists a Killing vector field $k^\mu$ which is timelike everywhere,
except perhaps in a compact closed region, such that ${\cal L}_k F_{\m\n}=
0$.  We can choose coordinates adapted to the Killing vector field such
that the metric is in ADM form:
\be
ds^2=-c^2 d\tau^2=-h(x)(dx^0-g_a(x) dx^a)^2+\gamma_{ab}(x)dx^a dx^b,\label{adm}
\ee
where $-h(x) = g_{00}$ and $g_{0a} = g_{0i}/h(x)$.
In the above, $(x)$ now denotes dependence on the spatial coordinates $x^a,
a,b,...=1,2,3$
only.
The electromagnetic 4-potential $A_\m(x^0,x^a)$ is of the form $(A_0(x),A_a(x))$,
i.e., it does not depend on the coordinate $x^0$.
In these coordinates, the Killing vector $k^\m=[1,0,0,0]$.
%
%

We now compute the functional derivatives of $S[x^0,x^a]$.  We fix the
coordinates $x^0, x^a$ at $\tau_1$ and $\tau_2$.  First, varying
$x^0$ only we obtain
\be
mc^2 h \omega={e\over c} A_0 +const.
\ee
We have written $\omega:=\dot x^0-g_a(x)\dot x^a$.  We restrict to observers
who measure time in the timelike Killing direction, that is, the unit
infinitesimal time interval is $\sqrt{h}(dx^0-g_a dx^a)$.  Hence, the
3-velocity is given by
\be
v^a:={c\over\sqrt{h}\omega}\dot x^a.
\ee
Writing $\beta^a:=v^a/c,\beta^2:=\gamma_{ab}\beta^a\beta^b$, we find
\be
\sqrt{h}\omega=(1-\beta^2)^{-1/2}.
\ee
If we now vary the space coordinate $x^b$, we get
\ba
 m \gamma_{ab}\left[\sqrt{1-\b^2}{d\over d\tau}\left({v^a\over\sqrt{1-\b^2}
}\right)+{1\over\sqrt{1-\b^2}}\lambda^a_{cd}v^c v^d\right]
&=&{mc^2\over\sqrt{1-\b^2}}\left[-\partial_b\left(\log{\sqrt{h}}\right)-\sqrt{h}
{v^a\over c}f_{ab}\right]\nn \\
&+&{e\over c}\left[{1\over\sqrt{h}}\partial_b A_0-{v^a\over c}\left(F_{ab}
+g_b\partial_a A_0-g_a\partial_b A_0\right)\right],
\ea
where $f_{ab}:=\partial_a g_b-\partial_b g_a$ and $F_{ab}:=\partial_a A_b
-\partial_b A_a$.  The quantities $\lambda^a_{bc}$ are the components of
the Christoffel connection for the 3D spatial metric $\gamma_{ab}(x)$.

The first term on the right hand side above is the Newtonian part of the
gravitational force; while the second term is the twist, or gravito-magnetic
part.  These terms can be defined covariantly as, respectively, the field
strengths associated with the
scalar potential $\phi:=\ln\sqrt{-g_{\m\n}k^\m k^\n}$ and the vector
potential $g_\m:=e^{-2\phi}
g_{\m\n}k^\n$.
We see that the twist field is trivial if the 1-form $\hat g:=g_\a dx^\a$ is
closed, that is $d\hat g=0$.  These 1-form fields are effectively three
dimensional.  That is, $H_\m=[0,\gamma_{cd}\epsilon^{abc}\partial_a g_b
/\sqrt{\gamma}]$.

The electric and magnetic fields can be expressed as:
\ba
E_a &:=& F_{0a}=-\partial_a A_0(x);  \\
B^a &:=& -{1\over2} \eta^{abc} F_{bc}=-{1\over2}\eta^{abc}
\left(\partial_bA_c(x)-
\partial_cA_b(x)\right).
\ea
In the above, $\eta^{abc}$ is the spatial LeviCivita {\it tensor}.
In addition to the usual electromagnetic gauge transformations
\be
A'_\m = A_\m - \f{\pa f}{\pa x^\m}~,~
\ee
one can also define `gravitational gauge transformation'
which preserves the stationarity of the metric, leaving
the equations of motion unchanged:
\be
x^a \rightarrow x^a~,~ x^0 \rightarrow x^0 + f(x^a),
\ee
under which:
\be
h \rightarrow h~,~ \c_{ab} \rightarrow \c_{ab}~,~
g_a \rightarrow g_a - \f{\pa f}{\pa x^a}~,~
{f}_{ab} \rightarrow {f}_{ab} ~.
\ee
%
%

We now set the charge $e=0$.  The uncharged massive particle will
now see an effective gravitational force with Newtonian and twist
components, given in 3D form as follows (in the $v/c <<1$ limit)
\footnote{ The following definitions have been used: $( {\vec a}
\times {\vec b} )_a = ~\eta_{abc} a^b b^c~, ( {\vec a} \times
{\vec b} )^a = {1\over2}\eta^{abc} a_b b_c~, ~(\vec \nabla \times
\vec a)^a = {1\over2} \eta^{abc} \le( \f{\pa a_c}{\pa x_b} -
\f{\pa a_b}{\pa x_c} \ri)~,~ \vec\nabla \cdot \vec a =
{1\over\sqrt{\c}} \pa_a \le( \sqrt{\c} a^a\ri)$ }:
\be
{\vec f}
= ~{mc^2}  \le[  - {\vec \nabla} \le( \ln \sqrt{h} \ri)
+ \f{ \sqrt{h}~{\vec v}}{c} \times
\le( {\vec\nabla} \times {\vec g} \ri) \ri].
\la{force4}
\ee

We see that motion in a stationary gravitational field
is analogous to that of a charged particle in an electromagnetic
field in a precise sense
and that the geodesic equation resembles the Lorentz force law.
The gravitational analogue of the electric charge is the energy
$mc^2$ of the particle.  The effective scalar and 3-vector potentials
are
\ba
\Psi&:=&\ln\sqrt{h},\nn \\
\vec{\cal A}&:=& \vec g.
\ea
Hence the effective 3-force is
\be
\vec f=mc^2 (\vec{\cal E}_g +\sqrt{h}\vec{\cal B}_g),
\ee
where $\vec{\cal E}_g:=-\vec{\nabla \Psi},\vec{\cal B}_g:=c\vec{\nabla}\times \vec g$.

\section{Landau Levels and Non-Commutativity in Gravity}
\la{sec3}

Now let us return to the motion of charged particle in an
electromagnetic field. The Lagrangian for the system is:
\be
L = \f{1}{2} m  {\dot {\vec x}}^2
+ \f{e}{c}~{\dot {\vec x}} \cdot {\vec A}  - e\Phi ~,
\la{l1}
\ee
where $A^\mu = [\Phi, {\vec A}]$ is the gauge potential.
Assuming that the charge
is in a constant magnetic field along the $x^3$-axis,
one can choose a symmetric gauge:
\be
{A}^i = -\f{B}{2}~\e^{ij} x^j  
\la{gauge1}
\ee
Now, if the magnetic field is very strong,
such that $eA/mc \gg v$, the characteristic velocity of the particle,
then as shown in \cite{djt} (see also \cite{jac2,magro}).
the kinetic term can be dropped from (\ref{l1}) and one obtains:
%
\be
L \rightarrow  \f{eB}{2c} \le[ -y {\dot x} + x {\dot y} \ri] - e\Phi~~,
\la{lag6}
\ee
Poisson brackets can be calculated using which the prescription
of \cite{fj}. Writing (\ref{lag6}) as
\be
L dt = a_i dx^i - e~\Phi dt
\ee
with $(x^1,x^2) = (x,y)$ and computing $f_{ij} \equiv a_{[j,i]}$, we get:
$f_{xy} = eB/2c$. Then:
\be
\{x, y\} = f_{yx}^{-1} = -\f{2c}{eB}~,
\ee
from which their commutator follows \cite{magro,jac2}:
\be
[x,y] = -i \f{2\hbar c}{eB} ~.
\la{nc6}
\ee
We note here that if we switch to polar coordinates $(\theta,r)$ in
space, then the commutator is
\be
[r,\theta]=-i \f{2\hbar c}{eB}{1\over r}.
\ee
Hence for very large $r$, spatial commutativity appears to be restored.

It is well known that the system under consideration has discrete
energy eigenvalues (`Landau levels'), which are given by:
\be
E_n = \f{eB\hbar}{mc} \le( n + \f{1}{2} \ri)~.
\la{ener6}
\ee
Thus, the above result also shows that for very strong magnetic fields,
the gap between ground and first excited state is large and the charge
is effectively confined to the lowest Landau level.

Motion in a Newtonian gravity field is described by the Lagrangian
Eq.(\ref{l1}) if
$e=mc^2$ and $\vec A=0$.  However, the presence of the factor $\sqrt{h}$
in the 3-force contribution from the twist potential prevents using that
Lagrangian to describe motion in the case when gravity has both a Newtonian
and a twist part.  In the remainder of this paper, we consider only the
case when there is no Newtonian limit (and no Coulomb electromagnetic
field).  In a the case of a curved 3-space, with metric $\gamma_{ab}(x)$,
we generalize Eq.(\ref{l1}) to
\be
L={1\over 2}m \gamma_{ab}\dot x^a \dot x^b +{e\over c} \dot x^a  A_a.
\ee
The charge $e$ is either an electric charge or in the case of
gravity, the energy $mc^2$, in which case $A_a$ is the twist part of
the gravitational field.

Now we see that an analogous effect takes place in the presence
of a strong stationary gravitational fields, if its magnetic analog
$ {\vec \nabla} \times {\vec g}$ is constant.  Consider first the
case of a flat 3-space metric $\gamma_{ab}=\delta_{ab}$.
The effective Lagrangian for the system is (\ref{l1}) with
$e/c=-2 m c$ and $\vec A=\vec g$.
A particle in this spacetime would be confined to the lowest energy level
given by (\ref{ener6}) with $\vec B \rightarrow \vec {\cal B}_g=c\vec\nabla\times\vec g$
a constant twist field.
Spatial non-commutativity follows as well from (\ref{nc6}).

Extension of the above results to charged particle confined to
$S^2$ or two dimensional hyperbolic plane ($H^2$) goes as follows
\cite{dunne2,comtet}: For $S^2$, the magnetic field is assumed to
be perpendicular to the spherical surface at every point and of
constant magnitude. The effective 2-space is a sphere of radius
$R$ with metric $R^2(d\theta^2+\sin^2{\theta})$.  The
corresponding monopole potential on $S^2$ is given by:
\be
A_\phi = 2 B R^2 \sin^2\le({\theta}/{2}\ri)~ 
~,~~
A_r = A_\theta =0~,~~
\la{gauge2}
\ee
where $B$ is the magnetic field and $R^2$, the radius of $S^2$
\footnote{Note that the potentials used here are
related to the Wu-Yang potential ($A_i'$) as $A_i = \sqrt{g_{ii}} {A'}^i$ }.
In this case, the
effective Lagrangian for strong magnetic fields is:
%
\be
L =  4 m c B R^2  ~\dot\phi \sin^2 \f{\theta}{2}~,
\la{efflag1}
\ee
from which it follows that $f_{\theta\phi} = -2 m c B R^2 \sin\theta$ and
%
\be
[ \theta, \phi ] = - \f{i \hbar }{2 m c B R^2 } \f{1}{\sin\theta}~.
\la{nctheta}
\ee
Note that, in this case the non-commutativity is not a constant.  In
fact, the non-commutativity is maximal near the poles $\theta=0,\pi$.
%
%
%
%
Similarly, for constant magnetic fields on a two dimensional hyperbolic
manifold $H_2$, $R^2 \rightarrow -R^2$, and the signature in the
RHS of (\ref{nctheta}) is reversed.
The energy levels in these cases are respectively:
\be
E_n =  \f{e B \hbar}{mc} \le( n + \f{1}{2} \ri)
\pm \f{\hbar^2}{2 m }~\f{n(n+1)}{R^2} ~.
\la{ener4}
\ee
For large $B$, the lowest Landau level effectively dominates the spectrum.
It is interesting to note however, that $B$ drops out of the second term.
The latter is an effect of the curvature of 3-space, and seems analogous
to the Casimir energy.
Eqs.(\ref{gauge2}) and (\ref{ener4}) reduce to (\ref{gauge1}) and
(\ref{ener6}) respectively in the limit $R^2 \rightarrow \infty$,
with the identification $\theta = r/R$ .

\section{Raychaudhuri-Som Geometries}
\label{sec4}

We now pose the following question: are there stationary spacetimes whose
`magnetic' parts $\vec {g}$ are of the same functional form as
(\ref{gauge1}) or (\ref{gauge2}), such that
particle in them would exhibit Landau levels?
Remarkably, the answer is in the positive. These spacetimes were discovered by
Som and Raychaudhuri in 1968 \cite{somray} as a class of solutions of
Einstein-Maxwell equations with a charged dust source.  These geometries have
the same
sort of causal pathologies as the G\"odel metric, and like the latter,
have sources which obey physically reasonable energy conditions. This
family of solutions was rediscovered later by one of us together with
A. Das \cite{gd}, but now in the form of a solution of
the Einstein-Maxwell equations
with a massive Klein-Gordon scalar field as source.

The metric in cylindrical coordinates is of the form
\be
ds^2 = -(c dt+\Omega r^2 d\phi)^2+dr^2+r^2 d\phi^2 +dz^2,\label{sr}
\la{somray1}
\ee
This solution has a charged
source which satisfies a Weyl-Majumdar-Papapetrou relation between the
charge $q$ and mass $\m$ of the source. In the units we use,
$\O= \m c/\hbar = qc/4{\sqrt{\p}}\hbar $.
Also, the Klein-Gordon field $\psi$ is chosen in a gauge so
that $\psi=\sqrt{2}$ and the non-zero components of the electromagnetic
potential are
\ba
A_t&=&{1\over c^3 \sqrt{2 G }},\\
A_\theta&=&{\Omega c^2 r^2 \over\sqrt{2 G }},
\ea
where $G$ is the Newton's constant.
The electric field and the `Newtonian part' of the gravitational field
are zero; the magnetic field and the `twist part' of the gravitational
field are constant and in the $z-$direction.  We see that these solutions
have one free parameter, which we can assign as the mass of the source
Klein-Gordon scalar field.

The metric (\ref{sr}) is homogeneous.
We will see below that the isometry group
of the metric is the direct product of translations in the $z-$direction with
the nilpotent Lie group, which has 4 parameters.
%
These geometries have recently reappeared in the literature, in the context
of string/M-theory. This is because they are special cases of a class
of geometries described by so-called G\"odel/AdS metrics of the form
\cite{rt,drukker}:
\be
ds^2 = - \le( cdt + 4\O R^2 \sin^2(\theta/2) d\phi \ri)^2
+ R^2 \le( d\theta^2 + \sin^2\theta d\phi^2\ri) +dz^2~.
\la{srdg2}
\ee
If $z=constant$, it can be shown that the above metric is a
solution of three dimensional Einstein-Maxwell equations with a positive
cosmological constant $1/R^2$ and a monopole potential of the form
$ A_0 = k;~A_\phi= 4k \O R^2 \sin^2(\theta/2)$, where
$k^2 = 1 + 1/(2\O R)^2$.
It is also a solution of the Einstein equations with more complicated matter field
sources in
four dimensions (\cite{rt}).
For $1/R^{2}= 2 \Omega^2, 1/R^{2} = 4 \Omega^2$, the metric becomes,
respectively, the `pure G\"odel', $AdS_3\times R$.
On the other hand, for $\O \rightarrow 0$ it
reduces to a metric on an $S^2$ of radius $R$, whereas for
$R \rightarrow \infty$ (and the identification $\theta = r/R$), it
reduces to the SR metric (\ref{somray1}). The corresponding
metric for two dimensional hyperbolic space $H_2$ can be obtained by the
substitution $R^2 \rightarrow - R^2$ \cite{oliveira}.

Also, we note that these geometries, together with appropriate
background fields, are exact backgrounds in string theory, in the sense that
a sigma model in the background is an exact conformal field theory
\cite{israel}.  Furthermore, there are black holes whose metrics are
asymptotically G\" odel/AdS or SR \cite{herdeiro,brecher}.

\section{Landau Levels and Non-Commutativity in SR Geometries}
\la{sec5}

It is remarkable that the `magnetic part' of the gravitational fields
of the stationary spacetimes described in the previous sections precisely match
that real magnetic fields which give rise to Landau levels in those
backgrounds.

\subsection{SR Spacetimes}

The SR metric (\ref{somray1}) in Cartesian coordinates is:
\be
ds^2 = -\left[dt-\Omega(y dx-xdy)\right]^2+dx^2 + dy^2 +dz^2~,
\ee
from which one can read-off the
effective potentials and fields:
%
%
implying
\be
h=1~~,~~\vec g = \Omega~\le(y,-x,0 \ri)
\ee
Note that ${\vec g}$ is precisely of the form of the vector potential
in Landau problem, Eq.(\ref{gauge1}). Furthermore, there are no electric
or Newtonian
gravitational fields.
Thus, the energy of the test-charge would be quantised (`gravitational
Landau levels') as:
\be
E_n = 2\Omega m  c\le( n + \f{1}{2} \ri)~~.
\la{ener1.2}
\ee
%
Further, if the gravito-magnetic fields are strong enough and the
charge is confined to the lowest Landau level, then from (\ref{nc6})
it follows that the spatial non-commutativity
experienced by the test particle is:
\be
[x,y] = -i {\hbar\over mc\Omega}
. \la{nc3}
\ee

\subsection{G\"odel Spacetimes}

Next, for the G\"odel like spacetimes (\ref{srdg2}) and its accompanying
gauge potential, the electric and magnetic parts can be read off:
\be
h = 1~~,~~{g}_{\phi} = - 4\O R^2 \sin^2(\theta/2)~,~g_r = g_\theta = 0 ,
\ee
for which:
\ba
\Psi &=& {\vec {\cal E}} = 0; \\
{\cal A}_\phi &=&
4\O \le( k -  \f{mc^2}{e}  \ri)
R^2 \sin^2\le( \f{\theta}{2} \ri)~.
\ea
The corresponding twist (gravito-magnetic) field is
\be
{\vec B} = 2 \O   {\hat R} .
\la{mag1.1}
\ee
The energy levels of the Landau-like system for an uncharged particle
of mass $m$ is
\be
E_n = 2 c \hbar \O
\le( n + \f{1}{2} \ri)
\pm \f{\hbar^2}{2 m}~\f{n(n+1)}{R^2} ~.
\la{ener4.1}
\ee
This results in a non-commutativity  from Eq.(\ref{nctheta}):
of
%
%
\be
[\theta,\phi] = - \f{i\hbar}{2 m c \O} \f{1}{\sin\theta}
\la{nctheta2}
\ee
Once again, the second terms in (\ref{mag1.1}), (\ref{ener4.1}) and
in the denominator of (\ref{nctheta2}) can be attributed to gravity
alone. Similarities between G\"odel and non-commutative spacetimes
were also noticed in \cite{romero}.

\section{Relation to the Fuzzy Sphere}

In this section, we show that the non-commutativity generated in the
context of a charged particle on $S^2$ with a monopole at the centre,
or equivalently for G\"odel spacetimes, is identical to the non-commutativity
on the fuzzy sphere. Let us consider Eq.(\ref{efflag1}) once again.
Using
\ba
\phi &=& \arctan \f{y}{x}~,~ \dot\phi= \f{x\dot y-y\dot x}{x^2+y^2}, \\
\sin^2\phi &=& \f{1}{2} \le( 1 - \f{z}{R} \ri)~,~x^2+y^2+z^2=R^2 .
\ea
Eq.(\ref{efflag1}) can be written as:
\be
L = \le( \f{BR^2e}{c} \ri) \le( \f{x\dot y - y\dot x}{x^2+y^2} \ri)
\le( 1- \f{\sqrt{R^2-(x^2+y^2)}}{R} \ri)~,
\ee
implying
\ba
Ldt &\equiv& a_x dx + a_y dy, \\
\mbox{where}~~~
a_x &=& \f{-y}{x^2+y^2}
\le( 1- \f{\sqrt{R^2-(x^2+y^2)}}{R} \ri)~
\f{BR^2e}{c}, \\
\mbox{   }~~~
a_y &=& \f{x}{x^2+y^2}
\le( 1- \f{\sqrt{R^2-(x^2+y^2)}}{R} \ri)~
\f{BR^2e}{c}~,
\ea
from which, one obtains:
\be
f_{xy} \equiv a_{[x,y]} = -\f{BRe}{c} \f{1}{z}~.
\ee
The above in turn implies the following Poisson and commutator brackets
\be
\{x,y\} = f_{yx}^{-1} = -\f{c}{BRe} z ~~,~
[x,y] = -\f{i\hbar c}{BRe} z~.
\ee
By changing the coordinate labels
$\{ x, y, z \} \rightarrow \{y,z,x\}$
and
$\{ x, y, z \} \rightarrow \{z,x,y\}$
respectively, the commutators $[y,z]$ and $[z,x]$ can be obtained
similarly, and the results can be summarised as:
\be
[x_i,x_j] = i \epsilon_{ijk} \lambda~x_k~,
\la{fuzzy1}
\ee
where $\lambda \equiv -\f{\hbar c}{BRe}$.
Note that Eq.(\ref{fuzzy1}) above is the algebra of coordinates
on a fuzzy sphere \cite{madore1}.
This implies that the Landau atom on $S^2$, or equivalently the
G\"odel universe can be regarded as concrete realisations of the fuzzy
sphere.

Finally, let us consider the stereographic projections of the fuzzy sphere
defined by:
\ba
y_{\pm} &=& 2 R x_{\pm} (R-z)^{-1} \equiv y_1 \pm iy_2,\\
%
\mbox{where} ~~~
x_{\pm} &=& x \pm i y~~
\ea
In the limit $R \rightarrow\infty$ and
$N \equiv 2R/\l \rightarrow \infty$, keeping $2R^2/N = \hbar c/eB$ fixed
($=\theta$ of ref.\cite{madore2}),
we get the non-commutative (fuzzy) plane defined by \cite{madore2} :
\be
[y_1,y_2] = - i \f{\hbar c}{eB},
\ee
which is identical to the commutation relation (\ref{nc6}), upto a multiplicative
factor of order unity.

\section{Emergence of the Star Product}

In this section, we follow the procedure outlined in \cite{jac3} to show
that the star product naturally emerges when one considers fluid flow
in the background of G\"odel-like spacetimes. We start with the Euler
equation for a non-relativistic fluid in a general curved spacetime
\cite{llfm} :
%
%
%
\be
\rho u^k u_{i;k} = p_{;i} - u_i \le( u^k p_{;k}\ri)~,
\la{euler1}
\ee
where $u^k= dx^k/ds$ is the $4$-velocity field and $\rho$ is the
mass density of the fluid. With the stationary metric (2) and as
before if we assume $h=1$ and $v/c << 1$ (non-relativistic
approximation) as well as $p_{,t}=0$ and $-\vec\nabla p/\rho=\vec
f$ as in \cite{jac3}, we obtain approximately:
\be
\f{\pa \vec v}{\pa t} + \le( \vec v\cdot \vec\nabla\ri) \vec v
= c~\vec v \times \le( \vec\nabla\times\vec g\ri) + \vec f~,
\la{eufin2}
\ee
which is identical to Eq(14) of \cite{jac3}, once when the latter
is scaled by $1/m$ and $(e/mc) \vec B$ is identified with
a {\it constant} gravito-magnetic field
$c \vec\nabla\times \vec g~.$ The constancy of
$\vec B$ ensures the identification with SR type metrics.

Further, from the covariant continuity equation
\be
\f{1}{\sqrt{-g}} \pa_i \le( \sqrt{-g} \rho u^i \ri) = 0~,
\ee
can be written in this case as
\be \pa_0 \le[ \sqrt{det(\gamma)h} \le\{ \rho h \le(
\f{1}{\sqrt{h}} + \f{\vec g \cdot \vec v}{c} \ri) \ri\} \ri] +
\pa_\alpha \le( \sqrt{det(\gamma) h}~ \f{\rho~
v^\a}{c\sqrt{1-v^2/c^2}} \ri)=0~. \ee
Again, for $h=1$ and $v/c << 1$ (such that the second term on the
left hand side drops out) and defining:
\be
\sqrt{det(\gamma)}~\rho \equiv \rho_1~,
\ee
we get:
\be
\dot\rho_1 + \vec\nabla\cdot \le( \rho_1 \vec v\ri) = 0~,
\la{conti2}
\ee
which is identical to Eq.(13) of \cite{jac3}. Consequently, the analysis
following Eq.(14) of \cite{jac3} go through and a star product emerges.
We sketch the steps in brief here.
We assume that the force $\vec f$ can be derived from a potential of the form:
\be
{\vec f} (r) = -\vec\nabla \f{\d}{\d \rho(r)} \int d\vec r~V~.
\ee
In the large gravito-magnetic limit,
Eqs.(\ref{eufin2}) and (\ref{conti2}) can be derived from the
Poisson brackets of $\rho$ and its canonical momentum $\pi$ with the Hamiltonian:
\be
H = \int d^2r \le( \rho~\f{\p^2}{2m} + V\ri)~,
\ee
with the following fundamental brackets
(and with the identification: $eb/mc \rightarrow 2\sqrt{8} \p c\m/\hbar$):
%
%
%
%
\ba
\le\{ \rho_1(\vec r), \rho_1(\vec r') \ri\} &=&
-\f{c}{eb} e^{ij} \pa_i \rho(\vec r)
\pa_i \d \le(\vec r - \vec r'\ri) \la{pb2}, \\
\{ \tilde \rho(\vec p), \tilde \rho(\vec q)\} &=&
-\f{c}{eb} e^{ij} p^i q^j \tilde\rho \le(\vec p + \vec q \ri)
\la{pb4},
\ea
where
\be
\tilde\rho (p) = \int d^2 r e^{i \vec p \cdot \vec r} \rho(\vec r) ~.
\ee
It can be shown that (\ref{pb2}) and (\ref{pb4}) is satisfied by
$\r(\vec r)$ of the form:
\be
\r_1(\vec r) = \sum_n \d \le(\vec r - \vec r_n\ri)~,
\ee
where $n$ labels the individual particles of the fluid, provided
their coordinates satisfy:
\be
\le\{ r_m^i, r_n^j \ri\} = \f{c}{eb}\e^{ij} \d_{mn}~.
\la{pb7}
\ee
Next, one quantises by writing the commutator bracket corresponding
to (\ref{pb7}) above:
\be
\le[ r_m^i, r_n^j \ri] = -i\hbar \f{c}{eb}\e^{ij} \d_{mn}
\la{commu6}
\ee
and assuming the following Weyl ordering:
\be
\tilde \rho (\vec p) = \sum_n e^{i \vec p \cdot \vec r_n}~.
\la{weyl1}
\ee
From (\ref{commu6}) and (\ref{weyl1}) and using the
Baker-Campbell-Hausdorff formula, one obtains:
\be
\le[ \tilde\r(p), \tilde \r (q) \ri]
= 2i \sin\le( \f{\hbar c}{2eb} e^{ij} p^i q^j \ri) \tilde\rho(p+q) ~.
\la{commuta8}
\ee
Finally defining:
\be
\langle f \ra = \int d^2 r \r(\vec r) f(\vec r)
= \f{1}{(2\p)^2}\int d^2 p~\tilde\rho(\vec p) \tilde f(-\vec p)~,
\ee
multiplying (\ref{commuta8}) by $\tilde f (-\vec p) \tilde g(-\vec q)$
and integrating gives:
\be
\le[ \langle f  \ra,\langle g \ra\ri] = \langle h \ra~,
\ee
with
\be
h(r) = \le( f\star g \ri) (\vec r) - \le( g \star f \ri) (\vec r) ~,
\ee
where
\be
 \le( f \star g \ri) (\vec r) \equiv
\exp( \f{i\hbar c}{2eb}\e^{ij} \pa_i \pa_j')
f(\vec r) g(\vec r') |_{\vec r' = \vec r}~.
\ee
Thus, we see how the star product emerges in the
context of G\"odel-like spacetimes, similar to its appearence
in the Landau atom.

\section{Conclusions}
\la{sec6}

In this article we have shown that the Newtonian and twist parts of
stationary metrics bear strong resemblance to ordinary
electric and magnetic fields.
The behaviour of particle geodesics in such backgrounds is similar to
that of charged particles in
electromagnetic fields. One important physical situation in
the case of the latter is the so-called Landau problem, in which
charges in a constant magnetic field have quantised energy levels. Morover,
if the field is very strong then the charge is effectively confined to the
lowest Landau levels. We observe that G\"odel like spacetimes
exhibit very similar behaviour for particle geodesics.
Furthermore, the effective non-commutativities that result in these
spacetimes are identical to that for the fuzzy sphere, under suitable
identification of parameters, and the star-product emerges when
one considers relativistc fluid motion in these spacetimes.
It would be interesting
to examine other physical implications of this result. To obtain the effective
fields due to electromagnetism and gravity,
we have made the simplifying
assumption of non-relativistic particle velocities. Generalisation
to arbitary velocities is expected to be straightforward and could provide
interesting corrections to our results. Similarly,
the $\theta$ dependence of the Poisson brackets in (\ref{nctheta})
merits further
investigation. Finally, it would be interesting to see whether the
non-commutativity studied here is related to non-commutativity in
string theory, where the Neveu-Schwarz $B$ field is taken to infinity
in the presence of $D$-branes \cite{sw}.
It is known that for ordinary field theories with quadratic actions, the
star-product reduces to the ordinary product \cite{sj}. Thus
quantities such as the two-point function should remain unchanged, not
acquiring corrections from the underlying non-commutativity.
We hope to report on these and related issues elsewhere.

\vs{.4cm}
\no
\ul{{\bf Acknowledgements:}}

\vs{.3cm}
\no
We thank A. Dasgupta, J. Madore and V. Husain for useful discussions.
SD thanks T. Sarkar and M. Walton for useful comments.
This work was supported by the Natural Sciences and Engineering Research
Council of Canada and funds of the University of Lethbridge.

\vs{.2cm}
\no

\end{document}